\documentclass{article}
\usepackage{spconf,amsmath,graphicx}
\usepackage{booktabs,tabularx,array,cite}
\usepackage[table]{xcolor}
\usepackage[hidelinks]{hyperref}
\usepackage{graphicx}
\usepackage{hyperref}
\usepackage{xurl}
\graphicspath{{figures/}}
\newcolumntype{Y}{>{\raggedright\arraybackslash}X}
\newcommand{\Fmicro}{F_{1,\mathrm{micro}}}
\newcommand{\mntd}{\mathrm{mNTD}}
\newcommand{\errtag}{\mathrm{Err}_{\mathrm{tagged}}}

\AtBeginDocument{%
  \setlength{\abovedisplayskip}{6pt plus 2pt minus 2pt}
  \setlength{\belowdisplayskip}{6pt plus 2pt minus 2pt}
  \setlength{\abovedisplayshortskip}{3pt plus 1pt minus 1pt}
  \setlength{\belowdisplayshortskip}{3pt plus 1pt minus 1pt}
}

\title{The NVVSpeech Challenge 2026: Overview and Lessons for Speech Understanding and Generation with Nonverbal Vocalizations}
\name{\parbox{0.96\textwidth}{\centering
Liumeng Xue$^{1,9}$, Yuang Cao$^{2}$, Jingbin Hu$^{2}$, Qirui Zhan$^{2}$,
Huakang Chen$^{2}$, Yibin Zheng$^{1}$, 
Yunxiang Chen$^{3}$, Houdun Liu$^{3}$, Su Feng$^{3}$, 
Haibin Wu$^{4}$, Zheng Lian$^{5}$, Wei Xue$^{6}$, 
Bengu Wu$^{7}$, Hui Bu$^{8}$, Lei Xie$^{2}$
}}

\address{\parbox{0.96\textwidth}{\centering
$^{1}$School of Intelligence Science and Technology, Nanjing University \\
$^{2}$Audio, Speech and Language Processing Group (ASLP@NPU), 
School of Computer Science, Northwestern Polytechnical University
$^{3}$Shenzhen Pimei Technology 
$^{4}$National Taiwan University
$^{5}$Tongji University 
$^{6}$Hong Kong University of Science and Technology 
$^{7}$Yutuzhineng 
$^{8}$AISHELL 
$^{9}$State Key Laboratory of Novel Software Technology, Nanjing University
}}
\begin{document}
% \ninept
\maketitle

\begin{abstract}

Nonverbal vocalizations (NVVs), such as laughter and sighs, convey meaning alongside words. NVV-aware speech systems must recognize and generate these events while preserving lexical accuracy and perceptual quality. Yet differences in annotation, resources, and evaluation complicate comparisons, and aggregate scores can obscure specific weaknesses. We review ten system descriptions from the NVVSpeech Challenge 2026, covering understanding and generation in Chinese and English across 16 event categories. Our analysis examines supervision construction, task adaptation, and inference-time or preference-based feedback. Results reveal persistent recognition difficulties with breath and gasp and trade-offs between event recognition and lexical accuracy, as well as between generation intelligibility and acoustic quality, with human listening providing complementary evaluation evidence. These findings identify priorities for advancing NVV-aware speech understanding, controllable generation, and comparable evaluation.
\end{abstract}


\begin{figure*}[!t]
    \centering
    \includegraphics[width=0.75\textwidth]{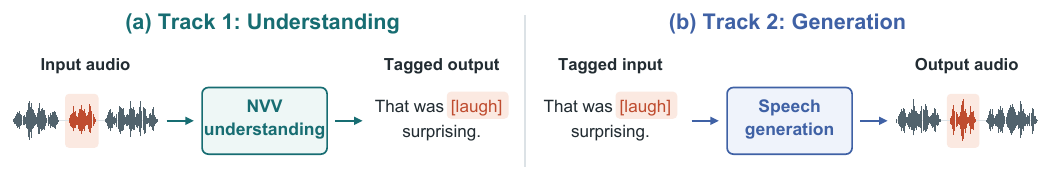}
    \vspace{-12pt}
    \caption{Overview of the two NVVSpeech Challenge tracks.
    Track~1 converts speech into a transcript with NVV categories
    and transcript-relative positions.
    Track~2 generates speech from a tagged transcript.
    Orange highlights the correspondence between a laughter event
    and its \texttt{[laugh]} tag. }
    % \vspace{-18pt}
    \label{fig:challenge_tracks}
\end{figure*}

 \vspace{-8pt}
\section{Introduction}
\label{sec:introduction}
 \vspace{-8pt}
Natural speech is a temporally organized communicative signal in which lexical content and nonverbal vocalizations (NVVs) jointly convey meaning. Laughter, sighs, breaths, and coughs may express affect, regulate interaction, or reflect physiological states. The central challenge is therefore not only to determine whether an NVV occurs, but also to identify its category, locate it relative to spoken words, and realize it with an appropriate acoustic form. A transcript can be lexically accurate yet omit important vocal behavior, while a synthetic utterance can remain intelligible yet fail to produce, control, or appropriately place a requested event~\cite{yang2026wesr,superbench}. NVV-aware speech technology must consequently address event recognition and generation together with lexical fidelity and perceptual quality.

Recent work has advanced this problem through complementary resources and methods. Annotated corpora support NVV-aware modeling in English, Mandarin, and bilingual settings~\cite{ye2025nonverbalspeech,
liao2025nvspeech,liao2026emilianv,mai2026mnv17,borisov2025nonverbaltts,
wu2025smiipnv}. Scalable annotation and synthetic-data pipelines further reduce the cost of constructing training data~\cite{ye2025nonverbalspeech,bai2026synparaspeech}. At the same time, evaluation has moved beyond aggregate transcription or audio-quality scores: WESR separates lexical errors from event detection and localization errors, whereas NVV-SuperBench evaluates controllability, placement, perceptual salience, and speech quality~\cite{yang2026wesr,
cao2026nvvlocator,superbench}. These efforts establish the main components of NVV research, but differences in taxonomies, annotation granularity, language coverage, and evaluation protocols still hinder direct comparison.

% The NVVSpeech Challenge 2026~\cite{challenge} brings these components into a shared bilingual setting for Mandarin Chinese and English. Its understanding and generation tracks use the same inventory of 16 NVV categories, inline-tag representation, held-out test sets, baselines, and scoring procedures. The open-resource protocol allows participants to construct training corpora and adapt pretrained models using disclosed external resources. It therefore supports comparison of supervision, adaptation, and inference strategies while making resource heterogeneity an explicit condition of interpretation.
% We review ten system descriptions, comprising seven understanding systems and three generation systems. The reported results show that additional data or class rebalancing does not necessarily resolve incomplete labels or limited acoustic diversity. Aggregate improvements may coexist with class-specific errors, reduced lexical accuracy, or larger language gaps, and auxiliary models may influence several stages of the pipeline, including annotation, candidate selection, and preference feedback. 

The NVVSpeech Challenge 2026\footnote{\url{https://nvvspeech-challenge.github.io/}} provides a shared Mandarin--English setting for NVV understanding and generation, with 16 event categories, inline tags, and standardized evaluation protocols. Its open-resource policy enables comparison of diverse training and adaptation strategies, with resource differences considered when interpreting results. We review ten systems, including seven for understanding and three for generation. The findings suggest that more data or class rebalancing alone may not overcome incomplete labels or limited acoustic diversity. Aggregate gains can also mask class-specific errors, lexical degradation, and language gaps, while auxiliary models shape annotation, candidate selection, and preference feedback.

Our contributions are therefore to: 
\textbf{(1)} We present the NVVSpeech Challenge 2026 as a common
bilingual platform for comparatively fair evaluation of NVV-aware
speech understanding and controllable generation under shared task
definitions, test sets, and scoring protocols.
\textbf{(2)} We systematically analyze ten systems from the perspectives
of supervision construction, model adaptation, and inference or
preference-based feedback, revealing how data, model, and feedback
designs affect NVV control and speech fidelity.
\textbf{(3)} We conduct a retrospective analysis of classwise,
language-specific, and within-study results, identifying persistent
breath and gasp errors, lexical--perceptual trade-offs, and
discrepancies between automatic and human evaluation, and derive
priorities for future NVV research.

\begin{table*}[!t]
\centering
\caption{
Overview of the 10 systems, comprising 8 challenge
system descriptions, 2 official baselines, and 2 non-participating
technical systems marked with \textsuperscript{*}. Reported
objective scores are given as Chinese / English / bilingual average
(ZH / EN / Avg.).
% \textsuperscript{*} Paper-only contribution instead of an official challenge entry.
% Track~1 scores are quoted from the system descriptions;
% Track~2 scores are from LALM-based evaluation and do not
% represent the final human-integrated ranking.
% Scores are not comparable across tracks.
% Bold indicates the highest ZH, EN, and Avg. scores within each track, including baselines and paper-only contributions.
}
\label{tab:systems}
\scriptsize
\setlength{\tabcolsep}{2.5pt}
\renewcommand{\arraystretch}{1.15}

\resizebox{\textwidth}{!}{%
\begin{tabular}{@{}lllllr@{}}
\toprule
Team / System & Backbone & Data / Supervision &
Training / adaptation & Auxiliary feedback / inference &
Scores (ZH / EN / Avg.) \\
\midrule

\rowcolor[RGB]{232,244,252}
\multicolumn{6}{c}{\textit{Track 1: Recognition}} \\

Moss-Transcribe & MOSS-Transcribe student &
Public/proprietary supervision &
Cold start; on-policy distillation &
Qwen3-Omni teacher &
66.39 / \textbf{66.24} / \textbf{66.32} \\

AcoustiX & Qwen3-ASR-1.7B &
Synthesis; voice conversion; insertion &
LoRA; class/language weighting &
Logit calibration &
\textbf{66.76} / 63.56 / 65.16 \\

XingChenAGI & Qwen3-ASR; LALM &
Cross-auditing; label refinement &
Curriculum; rebalancing; dual-stream tuning &
Teacher adjudication; HGC &
65.27 / 62.80 / 64.04 \\

SighLogger~\cite{sys166} & Qwen3-ASR-1.7B &
Label harmonization &
Square-root/uniform sampling; full tuning &
-- &
64.49 / 63.22 / 63.86 \\

SMIIP & MOSS-Audio; Whisper; Step-Audio &
Aligned NVV insertion &
Multi-backbone system &
Forced alignment; evidence fusion &
49.57 / 48.58 / 49.07 \\

Source-adaptive\textsuperscript{*}~\cite{sys157} & Whisper-medium &
Augmentation; media mining &
Full fine-tuning &
MLLM filtering; token reassignment &
55.45 / 51.76 / 53.61 \\

Audio Team\textsuperscript{*} & Qwen3-Omni MoE &
Synthetic NVVs; plain speech &
Routing-preserving adaptation &
Qwen3 forced alignment &
53.02 / 46.63 / 49.82 \\

Baseline & Whisper-medium &
NVV/ASR mixture &
Full fine-tuning &
NVV-token decoding &
34.05 / 32.59 / 33.32 \\

\midrule
\rowcolor[RGB]{232,244,252}
\multicolumn{6}{c}{\textit{Track 2: Generation}} \\

Tencent-HY~\cite{sys158} & NVV-aware DiTAR &
Targeted synthesis; long-tail data &
NVV tokens; SFT &
Multi-signal Best-of-5 &
76.22 / 69.66 / 72.94 \\

Moss-Transcribe & MOSS-TTS &
Review; event-site screening; 40/40/20 mix &
Two-stage SFT &
ASR-assisted filtering &
75.97 / \textbf{73.77} / 74.87 \\

Baseline & VoxCPM2 &
NVV-tagged speech pairs &
Backbone tuning; frozen VAE &
NVV-label mapping &
77.62 / 70.30 / 73.96 \\

Flow-DPO\textsuperscript{*}~\cite{sys163} & VoxCPM2 &
Chosen/rejected pairs &
RSFT; SFT-anchored Flow-DPO &
LALM preference labels &
\textbf{79.39} / 72.21 / \textbf{75.80} \\

\bottomrule
\end{tabular}%
}

\par\smallskip
% \begin{minipage}{\textwidth}
% \scriptsize
% \textsuperscript{*} Paper-only contribution; not an official challenge entry.
% \end{minipage}
\end{table*}
\vspace{-60pt}
 \vspace{45pt}
\section{Challenge Description}
\label{sec:challenge}
 \vspace{-8pt}

\subsection{Task Definition}
\label{sec:challenge_tasks}
 \vspace{-5pt}
The NVVSpeech Challenge covers Mandarin Chinese and English with
16 NVV categories: \textit{breath}, \textit{sniff}, \textit{laugh},
\textit{cry}, \textit{cough}, \textit{throat clearing}, \textit{sneeze},
\textit{sigh}, \textit{gasp}, \textit{snore}, \textit{yawn}, \textit{hum},
\textit{moan}, \textit{hiss}, \textit{lipsmack}, and \textit{burp}.
Inline tags such as \texttt{[laugh]} encode event identity and
transcript-relative position for two complementary tracks.
\textbf{Track~1: Understanding.}
Systems transcribe speech into words and NVV tags, preserving event
categories, and positions. Each prediction pairs an utterance ID
with its tagged transcript.
\textbf{Track~2: Generation.}
Systems synthesize tagged text, realizing the requested events while
preserving intelligibility, naturalness, and acoustic continuity.
% Each prompt requires one 16-kHz, mono, 16-bit PCM WAV file.

\vspace{-15pt}
\subsection{Data Resources}
\label{sec:challenge_data}
 \vspace{-10pt}
The open-resource setting provides no fixed training set. External data
and models are permitted with licensing compliance and disclosure. Meanwhile, outputs must be generated automatically.
 \textbf{Public resources.}
\href{https://huggingface.co/datasets/nonverbalspeech/nonverbalspeech38k}{NonVerbalSpeech-38K}~\cite{ye2025nonverbalspeech}
provides bilingual media speech, while
\href{https://huggingface.co/datasets/amphion/Emilia-NV}{Emilia-NV}~\cite{liao2026emilianv},
\href{https://huggingface.co/datasets/maimai11/MNV_17}{MNV-17}~\cite{mai2026mnv17},
and \href{https://huggingface.co/datasets/xunyi/SMIIP-NV}{SMIIP-NV}~\cite{wu2025smiipnv}
add word-level, performative, and expressive Mandarin data, respectively.
\href{https://huggingface.co/datasets/deepvk/NonverbalTTS}{NonverbalTTS}~\cite{borisov2025nonverbaltts}
provides aligned English NVVs and emotion labels;
\href{https://huggingface.co/datasets/shawnpi/SynParaSpeech}{SynParaSpeech}~\cite{bai2026synparaspeech}
adds synthetic speech; and
\href{https://huggingface.co/datasets/yfish/WESR-Bench}{WESR-Bench}~\cite{yang2026wesr}
annotates discrete and speech-overlapping events. The supplementary
Burp\_data\footnote{\url{https://huggingface.co/datasets/NVVSpeech-Challenge/Burp_data}}
contains 1,500 synthetic bilingual clips filtered by Gemini without human
verification~\cite{cao2026nvvlocator}.
Combining these resources requires label harmonization and checks of event
identity and placement.
\textbf{Evaluation sets.}
Track~1 releases 985 Chinese and 961 English recordings, withholding
reference transcripts and event annotations\footnote{\url{https://huggingface.co/datasets/NVVSpeech-Challenge/NVVSpeech-Challenge-Track1-Test-Set}}. Track~2 supplies 800 tagged
prompts per language\footnote{\url{https://huggingface.co/datasets/NVVSpeech-Challenge/NVVSpeech-Challenge-Track2-Test-Set}}.

\vspace{-15pt}
\subsection{Evaluation Metrics}
\label{sec:challenge_evaluation}
 \vspace{-3pt}
For language $\ell$, the objective scores of Track 1 and Track 2 are computed as:
{\small
\begin{equation}
\begin{aligned}
S_1(\ell)=100[&0.70\Fmicro+0.20(1-\mntd)\\
             &+0.10(1-\min(\errtag,1))],\\
S_2(\ell)=100[&0.30A+0.25P+0.15N+0.15Q+0.15E].
\end{aligned}
\label{eq:challenge_scores}
\end{equation}
}
For each track, the final leaderboard score is the arithmetic mean
of the Chinese and English scores. 
For \textbf{Track 1: Understanding}, $\Fmicro$ pools events across classes,
requiring matching categories and aligned positions. Multi-event normalized
tag distance ($\mntd$) measures placement error; $\errtag$ is Chinese CER
or English WER on the tagged transcript, treating each tag as one token.
For \textbf{Track~2: Generation}, evaluation follows the framework of
\href{https://github.com/lmxue/NVV-SuperBench}{NVV-SuperBench}~\cite{superbench}.
Normalized components measure event type, count, and position ($A$),
audibility and recognizability ($P$), naturalness and continuity ($N$),
audio quality ($Q$), and contextual expression ($E$). Missing or nearly
inaudible target events receive zero for $A$ and $P$.
Both tracks equally average Chinese and English scores. Track~2 uses LALM
assessment for the public leaderboard and combines verified LALM scores
$O$ with human listening scores $H$ for final ranking:
$S_{2,\mathrm{final}}=0.30O+0.70H$, where $O$ and $H$ are bilingual
averages on a 0--100 scale. 

\vspace{-10pt}
\subsection{Baseline Systems}
\label{sec:challenge_baselines}
 \vspace{-5pt}
\textbf{Track~1: Understanding.}
The Track~1 baseline\footnote{\url{https://github.com/NVVSpeech-Challenge/NVVSpeech-Challenge-Track1-Baseline}}
fully fine-tunes Whisper-medium~\cite{nvvbaseline_whisper}, reassigning
16 token IDs to NVV tags without expanding the vocabulary. Training mixes
NVV-annotated and conventional ASR speech at a 60/40 ratio; greedy decoding
jointly produces words and tags without timestamps.
\textbf{Track~2: Generation.}
The Track~2 baseline\footnote{\url{https://github.com/NVVSpeech-Challenge/NVVSpeech-Challenge-Track2-Baseline}}
fine-tunes the VoxCPM2 backbone~\cite{zhou2026voxcpm2} on paired audio and
tagged text while freezing its audio VAE. English-form tags are mapped to
Chinese training labels without changing lexical text.
% native 48-kHz outputs require resampling for submission.

\vspace{-6pt}
\section{Participating and Related Approaches}
\label{sec:approaches}
\vspace{-8pt}

Table~\ref{tab:systems} summarizes participating systems and paper-only
contributions. Their designs address three distinct bottlenecks:
the reliability and coverage of event supervision, the adaptation of
pretrained models to tagged speech, and the use of auxiliary feedback.
Across these stages, a common design objective is to introduce NVV
capabilities while preserving the surrounding speech. The distinctive
approaches differ in where they intervene: labels, token distributions,
expert routing, or generated candidates.

\vspace{-12pt}
\subsection{Backbone }
\label{sec:backbones}
 \vspace{-8pt}
Understanding systems combine ASR backbones with broader audio-language
models. Qwen3-ASR underlies AcoustiX, XingChenAGI, and SighLogger;
Whisper-medium supports the baseline and source-adaptive study~\cite{sys157,sys166}.
SMIIP combines MOSS-Audio, Whisper, and Step-Audio, while Audio Team
adapts Qwen3-Omni's MoE decoder. Moss-Transcribe separates the training
teacher, Qwen3-Omni, from the deployed MOSS-Transcribe student.
Generation uses DiTAR, MOSS-TTS, and VoxCPM2, with VoxCPM2 shared by
the baseline and Flow-DPO~\cite{sys158,sys163}. Backbone choice therefore
determines the available adaptation mechanisms: token-level transfer
requires compatible token spaces, routing adaptation requires accessible
experts, and flow-based generation permits flow-matching preference
surrogates. Conversely, the diversity of strategies built on Qwen3-ASR
shows that choosing a backbone leaves substantial data and optimization
decisions unresolved. The heterogeneous training and inference settings
preclude attributing score differences to architecture alone.

\vspace{-12pt}
\subsection{Supervision and Targeted Coverage}
\label{sec:supervision}
 \vspace{-8pt}
% \noindent\textbf{Separate label quality from training exposure.}
SMIIP audits annotations, while the source-adaptive study~\cite{sys157}
filters public data and mines spontaneous NVVs. Co-Refine takes a more
structured approach: heterogeneous models trained on disjoint partitions
cross-audit weak labels, with teacher adjudication and human review for
disagreements. Its distinction is to make label correction an explicit
stage rather than treating existing tags as fixed targets. Rebalancing
addresses a separate problem: SighLogger adjusts class exposure through
square-root sampling~\cite{sys166}, whereas XingChenAGI uses curriculum
learning and clipped resampling. Repetition changes exposure but cannot
correct a mislabeled event.

% \noindent\textbf{Augment the missing supervision.}
Aligned insertion supplies event-position targets; tagged synthesis and
voice conversion expand contexts and speakers, as in AcoustiX.
For generation, MOSS-TTS screens the requested event region as well as
lexical completeness and audio quality. This local focus matters because
a fluent utterance may still omit its requested NVV. ASR provides
supporting evidence, rather than a complete test of event presence or
category correctness.

\begin{figure*}[t]
    \centering
    \begin{minipage}[t]{0.32\textwidth}
        \vspace{0pt}
        \centering
        \includegraphics[width=\linewidth]{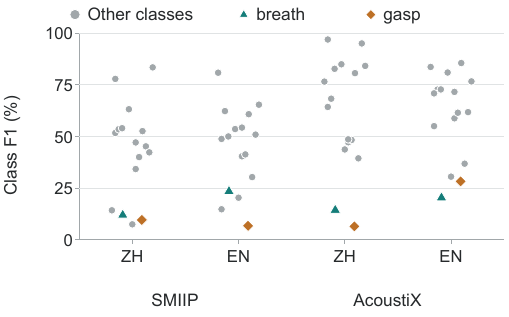}
        \par\smallskip
        \vspace{-5pt}
        {\small (a) Class-wise F1\par}
    \end{minipage}\hfill
    \begin{minipage}[t]{0.32\textwidth}
        \vspace{0pt}
        \centering
        \includegraphics[width=\linewidth]{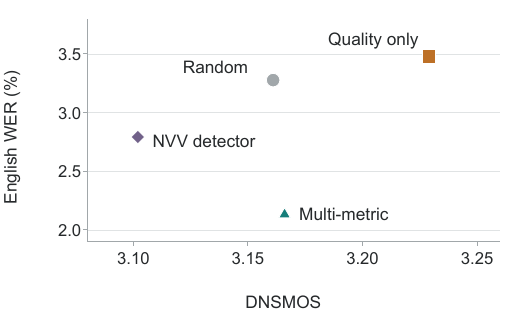}
        \par\smallskip
        \vspace{-5pt}
        {\small (b) Candidate selection\par}
    \end{minipage}\hfill
    \begin{minipage}[t]{0.32\textwidth}
        \vspace{0pt}
        \centering
        \includegraphics[width=\linewidth]{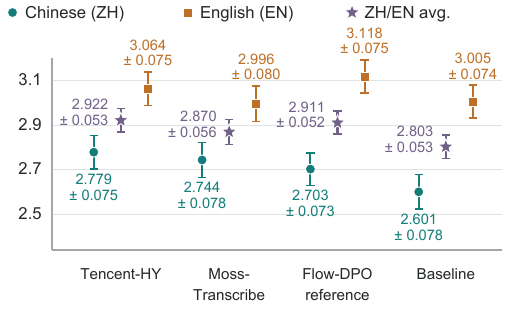}
        \par\smallskip
        \vspace{-5pt}
        {\small (c) Weighted listening scores\par}
    \end{minipage}
    \vspace{-5pt}
\caption{
Reported results: (a) classwise NVV F1 for SMIIP
and AcoustiX; (b) Tencent-HY candidate selection
on a shared five-candidate pool; and
(c) weighted human listening scores.
ZH/EN denote Chinese/English; stars indicate
bilingual means.
}
 %   \caption{%
 %   (a) F1 scores for all 16 classes reported by SMIIP and AcoustiX;     breath and gasp rank among the four lowest in every setting.     (b) Tencent-HY's DiTAR selection rules on the same five-candidate     pool; higher DNSMOS and lower WER are preferred.     (c) Track~2 listening scores for Chinese (ZH) and English (EN).     Stars denote bilingual averages.}
    \label{fig:results_analysis}
    
    % \vspace{-16pt}
    % \caption{Results and retrospective analyses.
    % (a) All 16 class F1 values reported by SMIIP and AcoustiX.
    % Each marker represents one class, with horizontal offsets
    % for clarity; breath and gasp rank among the four lowest
    % in every setting.
    % (b) Tencent-HY's DiTAR selection rules evaluated on the
    % same five-candidate pool using local objective metrics;
    % higher DNSMOS and lower WER are preferred.
    % (c) Track~2 weighted listening-test results for Chinese
    % (ZH) and English (EN). Stars indicate the bilingual
    % average; labels and error bars report mean $\pm$
    % uncertainty. Flow-DPO is a non-participating technical
    % reference.}

\end{figure*}
\vspace{-12pt}
\subsection{Adapting and Preserving Pretrained Structure}
\label{sec:adaptation}
 \vspace{-8pt}
% \noindent\textbf{On-policy transfer and routing preservation.}
Moss-Transcribe's tokenizer-aligned distillation is distinctive: after
supervised initialization, the student generates prefixes on official
unlabeled audio, and a frozen Qwen3-Omni teacher scores those same
prefixes for reverse-KL training. Unlike supervision restricted to
reference prefixes, this places feedback on trajectories the student
actually visits; the shared tokenizer enables direct token-level
comparison. Audio Team intervenes inside the model instead: it disables
MoE load balancing and adds Prior Intervention Routing over a frozen
gate, aiming to adapt event behavior while retaining pretrained expert
specialization. These designs separate transferring output behavior from
preserving internal computation; their aggregate scores do not isolate
either mechanism's contribution.

% \noindent\textbf{Represent events without terminating speech.}
DiTAR adds NVV tokens and modifies stop prediction so that internal
vocal events are distinguished from utterance endings~\cite{sys158}.
This targets a structural failure mode beyond class imbalance.
MOSS-TTS instead emphasizes two-stage fine-tuning on curated mixtures,
adapting through supervision and training schedule.

\vspace{-12pt}
\subsection{Feedback at Inference and During Training}
\label{sec:feedback}
 \vspace{-8pt}
Feedback has two roles: revising or selecting current outputs, and
changing the model that produces future outputs. For understanding,
SMIIP fuses event evidence around a stable transcript, while Co-Refine's
hypothesis-guided correction combines the audio with an initial
task-specific prediction. For generation, Tencent-HY selects among five
candidates using complementary event, ASR, LALM, and quality signals~\cite{sys158}.
Flow-DPO instead incorporates LALM preferences into VoxCPM2 training~\cite{sys163}:
an utterance-level flow-matching surrogate supplies pairwise supervision,
while a chosen-sample SFT anchor retains direct fitting to preferred
outputs. The anchor complements relative preference learning with an
absolute training target. Candidate selection spends computation at
inference; preference optimization moves feedback into model updates.
Both require validation of the feedback signals against event control,
lexical fidelity, and perceived quality.

\vspace{-13pt}
\section{Results and Retrospective Analysis}
\label{sec:results}
 \vspace{-8pt}
Table~\ref{tab:systems} and Fig.~\ref{fig:results_analysis} provide
complementary views of system performance: aggregate scores, classwise
errors, candidate-selection trade-offs, and human listening results.
Moss-Transcribe has the highest reported Track~1 bilingual score (66.32)
and leads participating Track~2 systems under LALM evaluation (74.8734).
Flow-DPO~\cite{sys163} scores 75.7984 but is a non-participating reference.
The following analysis preserves each study's evaluation scope;
cross-system differences do not isolate individual design contributions.

% The results in Table~\ref{tab:systems} summarize the objective evaluation results for both tracks. For Track~1, the three highest reported bilingual scores are obtained by Moss-Transcribe (66.32), AcoustiX (65.16), and XingChenAGI (64.04). For Track~2, among official participating systems, Moss-Transcribe ranks first under LALM evaluation (74.8734), followed by the baseline (73.9602) and Tencent-HY (72.9406). Flow-DPO achieves a higher LALM score (75.7984), but it is a paper-only, non-participating reference system. After combining objective and subjective evaluations as shown in Fig.~\ref{fig:results_analysis}(c), Tencent-HY obtains the highest final score among participating systems (55.5123). The following analysis examines the design choices and evaluation patterns behind these results.

% The Track~2 evaluations in Table~\ref{tab:systems}
% place Moss-Transcribe first among participating systems under LALM
% evaluation (74.8734). Flow-DPO~\cite{sys163} scores 75.7984 but serves
% only as a non-participating technical reference. Combining objective
% scores with the subjective listening results in
% Fig.~\ref{fig:results_analysis}(b) places Tencent-HY first among
% participants (55.5123). The analysis below combines these 
% evaluations with findings. 

\vspace{-13pt}
\subsection{Class Weaknesses and Rebalancing}
\label{sec:reliability}
 \vspace{-8pt}
Fig.~\ref{fig:results_analysis} (a) reveals a limitation obscured by
aggregate scores: breath and gasp rank among the four weakest classes
for both SMIIP and AcoustiX in both languages. AcoustiX's higher overall
score therefore coexists with persistent weaknesses in these events.
This recurrence identifies priorities for targeted error analysis,
but does not establish whether rarity, annotation ambiguity, acoustic
salience, or placement errors are responsible.

Rebalancing should consequently be validated rather than prescribed
from class frequency alone. In SighLogger's separate local comparison,
square-root sampling scores 66.25, versus 62.33 for natural and 63.51
for uniform sampling~\cite{sys166}. Moderate rebalancing helps in this
setting, but stronger balancing is not uniformly better. Together,
these observations motivate distinguishing omissions, category
confusions, and placement errors before choosing augmentation or
sampling interventions.

\vspace{-10pt}
\subsection{NVV Control and Speech Fidelity}
\label{sec:lexical}
 \vspace{-8pt}
Fig.~\ref{fig:results_analysis}(b) isolates the effect of candidate
selection within Tencent-HY's fixed five-candidate pool~\cite{sys158}.
DNSMOS-only selection maximizes DNSMOS among the compared rules, but
multi-metric selection lowers English WER from 3.480\% to 2.132\%
with lower DNSMOS. Thus, a quality-oriented selector need not choose
the most lexically faithful output. This figure supports a trade-off
between the plotted metrics; it does not by itself establish which
selector best realizes NVV instructions or human preferences.

A related tension appears during recognition adaptation: the
source-adaptive study increases English event F1 from 0.2964 to 0.5081
while text-only WER rises from 25.58\% to 34.98\%~\cite{sys157}.
These within-study comparisons motivate explicit checks of lexical
fidelity alongside event control and acoustic quality, during both
training and output selection.

\vspace{-10pt}
\subsection{Objective and Subjective Evaluation}
\label{sec:languages}
 \vspace{-8pt}
Automatic evaluation and human listening do not yield identical
system rankings. As shown in Table~\ref{tab:systems} and
Fig.~\ref{fig:results_analysis}(c), Moss-Transcribe achieves the
highest LALM score among participating generation systems, whereas
Tencent-HY ranks first after human listening scores are incorporated
with a 70\% weight. This change indicates that the automatic ranking
does not fully capture the preferences reflected in the listening
evaluation, although it does not establish statistically significant
perceptual differences between systems.

This discrepancy is particularly relevant when auxiliary models
also guide system development. Tencent-HY uses model-based signals
for candidate selection, while Flow-DPO uses LALM preferences for
training~\cite{sys158,sys163}. Improvements under automatic assessment
should therefore be checked against human judgments, especially when
optimization and evaluation rely on similar feedback signals.
These findings highlight the need to complement automatic evaluation
with human listening tests, since higher automatic scores do not
necessarily indicate better perceived speech quality.

\vspace{-10pt}
\section{Conclusion}
\label{sec:conclusion}
 \vspace{-10pt}
The NVVSpeech Challenge provides a shared Mandarin--English benchmark for event-aware understanding and controllable generation. Our review of ten system descriptions shows that supervision, adaptation, and feedback design shape performance: breath and gasp remain difficult, event-aware gains may reduce lexical accuracy, and candidate selection trades intelligibility against acoustic quality. Human listening offers a complementary perspective on system quality. Future work should improve annotation, preserve lexical and perceptual quality, and report resource and feedback conditions alongside multidimensional evaluation.

\bibliographystyle{IEEEbib}
\bibliography{main}

\end{document}